# A Systematic Mapping Study on Security in Agile Requirements Engineering

Hugo Villamizar
Software Engineering Laboratory
Pontifical Catholic University of
Rio de Janeiro (PUC-Rio)
Rio de Janeiro, Brazil
hvillamizar@inf.puc-rio.br

Marcos Kalinowski
Software Engineering Laboratory
Pontifical Catholic University of
Rio de Janeiro (PUC-Rio)
Rio de Janeiro, Brazil
kalinowski@inf.puc-rio.br

Marx Viana
Software Engineering Laboratory
Pontifical Catholic University of
Rio de Janeiro (PUC-Rio)
Rio de Janeiro, Brazil
mleles@inf.puc-rio.br

Daniel Méndez Fernández
Software & Systems Engineering
Technical University of Munich
(TUM)
Garching, Germany
daniel.mendez@tum.de

*Abstract*—[Background] The rapidly changing business environments in which many companies operate is challenging traditional Requirements Engineering (RE) approaches. This gave rise to agile approaches for RE. Security, at the same time, is an essential non-functional requirement that still tends to be difficult to address in agile development contexts. Given the fuzzy notion of "agile" in context of RE and the difficulties of appropriately handling security requirements, the overall understanding of how to handle security requirements in agile RE is still vague. [Objective] Our goal is to characterize the publication landscape of approaches that handle security requirements in agile software projects. [Method] We conducted a systematic mapping to outline relevant work and contemporary gaps for future research. [Results] In total, we identified 21 studies that met our inclusion criteria, dated from 2005 to 2017. We found that the approaches typically involve modifying agile methods, introducing new artifacts (e.g., extending the concept of user story to abuser story), or introducing guidelines to handle security issues. We also identified limitations of using these approaches related to environment, people, effort and resources. [Conclusion] Our analysis suggests that more effort needs to be invested into empirically evaluating the existing approaches and that there is an avenue for future research in the direction of mitigating the identified limitations.

I. INTRODUCTION

Security is a critical non-functional requirement and needs consideration as an integral part of software development methodologies. However, developing secure software is not a trivial task as it requires to address security from early development stages throughout the whole software lifecycle [16]. Yet, in the majority of software projects, security is often dealt with in retrospective when the system has already been designed and put into operation [19].

Traditional, document-intensive requirement engineering includes practices to elicit, analyze, specify, and validate and verify requirements [9]. In such practices, the communication of requirements is mostly based on formal and extensive documentations [1]. Although the usage of such traditional approaches is widespread, especially in the software industries dealing with security and safety critical systems, the necessity of delivering projects in shorter periods has triggered the desire and need to change towards gradually increasing the usage of agile methods. In response to this, agile Requirements Engineering (RE) has experienced a rise. Agile RE has already helped to address some specific problems of requirements engineering, but it also has caused others [20]. Typical characteristics of agile RE practices include narrowing down the focus on requirements specification activities in a human-centric manner (e.g., by using user stories) while explicitly aligning with frequently changing needs (e.g., via short, iterative feedback cycles and product backlogs) [25].

However, an emerging challenge comes by that the underlying philosophies of agile software (requirements) practices and of secure development seem to be opposing each other [2][3]. This renders a common understanding of how to integrate security requirements with agile practices difficult. First valuable solution proposals have been contributed in this direction, but while the research communities have produced secondary studies to structure the publication landscape of the areas of agile requirements [6][11][12][18] and security [13][19], each in isolation, we still lack a common and integrated view on the integration of security requirements in agile practices.

In response to this challenge, we contribute a first step in synthesizing existing work on integrating security requirements in agile practices via a secondary study. In particular, we report on a systematic mapping study with the research objective to **understand what are the approaches that have been proposed to handle security requirements in agile software projects**. We aim at characterizing these approaches in terms of the following research questions:

- **RQ1.** In the context of which agile methods have the approaches been proposed or applied?
- **RQ2.** Which RE phases do the approaches address?
- **RQ3.** How are security requirements handled in each approach?
- **RQ4.** What is the research type facets of the approaches?
- **RQ5.** Which kind of empirical evaluations have been performed?
- **RQ6.** What are the limitations faced by the identified approaches?

While the first questions 1-3 aim at structuring the publication landscape in a conceptual manner, the last two shall shed light on the current state of reported evidence. We found that Scrum is the main agile method explored by the researchers. Several artifacts and activities (e.g., abuser story and vulnerability analysis) have been proposed to address security requirement specification aspects. Further, 12 out of 21 papers (57%) refer to new solution proposals. In that context, we outline worthwhile fields of future investigation, e.g. specialized approaches to validate security requirements in agile contexts, and conclude by postulating that there is a need for further evaluation research to test the sensitivity of the solution proposals in practical context. Especially the latter serves an effective dissemination of new approaches into practice.

The remainder of this paper as organized as follows. Section II provides the background and an overview on related work. Section III describes the mapping study protocol and how it was applied. Section IV presents the mapping study results. Section V discusses the results. Section VI describes the threats to the validity of our study. Finally, Section VII presents the concluding remarks.

## II. BACKGROUND AND RELATED WORK

This section provides the background on agile requirements and security requirements and describes the related work, including other secondary studies.

### A. Agile RE

The term "agile requirements engineering" emerged in response to the agile manifesto in 2001. It is used to define the ''agile way'' of planning, executing and reasoning about RE activities [12]. Yet, not much is known about the challenges posed by the collaboration-oriented agile way of dealing with RE activities. Ramesh *et al.* [23] performed a multi-case study with 16 organizations that develop software using agile methods. They identified that agile RE practices resulted in challenges regarding neglected non-functional requirements, minimum documentation and no requirements verifications [23]. The recent report from the Naming the Pain in Requirements Engineering (NaPiRE) initiative [20] extends the list of challenges with, inter alia, communication flaws between project teams and the customer and underspecified requirements that remain too abstract and, thus, are not measurable. This already gives a picture on the difficulties of dealing with non-functional requirements in agile environments and it is reasonable to believe that security requirements are no different in this respect.

### B. Security Requirements (SR)

Software development should be conducted with security in mind at all stages and it should not be an afterthought [16]. SR have traditionally been considered "quality" requirements [5][7]. Like other quality requirements, they tend not to have simple *yes/no* satisfaction criteria (e.g., as acceptance criteria). Determining whether a quality requirement has been satisfied (well enough) is a complex task, especially when it comes to SR given their technological and socio-economic contexts and implications. Haley *et al.* [10] presents some challenges related to SR. First, people generally think about and express SR in terms of "bad things" (negative properties) to be prevented. It is very difficult, if not impossible, to measure negative properties. Second, for SR, the tolerance on "satisfied enough" is small, often zero, given the implications of non-compliance, and stakeholders tend to want SR satisfaction to be very close to *yes*. Third, the effort stakeholders might be willing to dedicate to satisfying SR also depends on the likelihood and impact of a failure to comply with SR. However, while we cannot justify expenses that are not in tune with cost-benefit calculations, it is exactly this what we often cannot estimate upfront when it comes to security [10].

First work in response to the challenges in dealing with SR considers the use of quality models to reuse SR as they can indicate the social and technological implications of such requirements early on (see, e.g., [17]). However, we still have a limited understanding of how to integrate SR to overcome the challenges. While this is already evident in traditional plan-driven software development projects, we even know less about the integration of security into agile software development.

### C. Related work

To the best of our knowledge, we are not aware of mapping studies concerning security in the context of agile RE. However, we identified papers that present secondary studies concerning agile RE and SR, each in isolation. The remainder of this subsection summarizes these studies and their findings.

Inayat *et al.* [12] conducted a systematic literature review on agile RE research published between 2002 and mid of 2013. In this study, 21 papers were identified and the review identified 17 practices of agile RE, 5 challenges traceable to traditional RE that are resolved by agile RE and 8 practical challenges posed by the practice of agile RE. Their findings suggest that agile RE research needs additional attention and that more empirical research is required to better understand the impact of agile RE practices [12]. With respect to security, this study only states that the use of agile RE opens up challenges for the software industry, including the lack of approaches to deal with non-functional requirements such as security and scalability.

Medeiros *et al.* [18] conducted a mapping study on RE in agile projects based on evidence from industry. In this study, they identified 24 papers published before 2014. The goal of this study was to conduct an industrial exploratory study to investigate how RE is used in projects that adopt agile methodologies. Their findings show that low user involvement and constant change of requirements were identified as the main challenges to be overcome [18]. Their study did not emphasize particular types of non-functional requirements, such as SR.

Heikkila *et al.* [11] conducted a mapping study on agile RE in September 2014. In this study, 28 papers were identified and analyzed. Their results illustrate benefits and problematic areas of agile RE. The authors also reported solutions for the identified problems [11]. Security, however, was not in scope in their study.

More recently, Curcio, *et al.* [6] also conducted a SM study on RE in agile contexts. They identified 104 papers published until 2017. According to their findings, agile methodologies have not adequately modeled non-functional requirements and their potential solutions. They conclude that aspects related to maintainability, portability, security or performance are often neglected due to the tendency of using business value as the main prioritization criteria [6], reinforcing the difficulties of estimating costs and benefits of non-functional requirements upfront (see also our concluding discussion in Sect. B).

Regarding SR, Mellado *et al*. [19] and Khan and Ikram [13] conducted a systematic reviews on this topic. The authors identified many techniques, processes and methodologies. However, both studies did not focus on agility and most of the approaches are not well aligned with the agile development philosophy.

To the best of our knowledge there are no secondary studies specifically structuring the publication landscape of SR in the context of agile practices. Moreover, we could observe valuable solution proposals (e.g., [P2][P20][P21]) that are not covered in the existing secondary studies. The mapping study presented hereafter addresses this relevant gap.

III. SYSTEMATIC MAPPING PROTOCOL

Systematic literature studies have become a valuable means to structure reported knowledge in a systematic manner. Systematic Mapping (SM) studies constitute one form of secondary study that allows to broadly categorize relevant solutions and concepts in a specific field and visualize the coverage and maturity of an entire research field [15] through a systematic procedure whose purpose is to identify the extent and nature of the primary studies available in the area [4][22]. The main reasons to perform a SM are to systematically identify gaps in the current body of research and support the planning of new research, avoiding the unnecessary duplication of effort and error [4].

The SM study was performed following the guidelines proposed by Kitchenham and Charters [14] and the SM-specific guidelines proposed by Petersen *et al*.[22]. After identifying the need for the review (cf. Section II), we define the research questions, search strategy and inclusion/exclusion criteria.

*A. Research Objectives and Questions*

The main research objective is to **understand what are the approaches that have been proposed to handle SR in software agile projects**. The following research questions were derived from the objective in order to further characterize the identified approaches.

**RQ1. In the context of which agile methods have the approaches been proposed or applied?** This question aims at getting a general overview and understanding the agile methods (e.g., Scrum, XP) for which SR approaches have been proposed.

**RQ2. Which RE phases do the approaches address?** The aim of this question is to identify the specific RE phase (following the ReqMan [1] process framework classification according to elicitation, analysis, specification, verification & validation) that each identified approach explicitly addresses. While it is not our intention to analyze particular RE practices,

[1] http://www.reqman.de

this categorization helps understanding the purpose of the introduced approaches.

**RQ3. How are SR handled in each approach?** This question involves providing a short description of the approaches and analyzing how each approach suggests to address SR, i.e., how solution options are typically built.

**RQ4. What is the research type facets of the approaches?** The purpose of this question is to classify the papers according to its research type facet. We adopt the classification scheme proposed by Wieringa *et al.* [26], including the following categories: evaluation research, solution proposal, validation research, philosophical paper, opinion paper, and experience paper.

**RQ5. Which kind of empirical evaluations have been performed?** The purpose of this question is to identify what types of empirical studies have been used to assess the approaches, thus, focusing on the research type facets evaluation and validation research from the previous question. Obtaining this information allows us to get a first idea of the state of reported evidence in the field.

**RQ6. What are the limitations faced by the identified approaches?** This question aims at outlining the limitations reported for the approaches within the identified papers.

While the first three questions shall structure the publication landscape, the following three questions shall provide a foundation to discuss the current state of evidence and implications on future research.

*B. Search Strategy*

The mapping study employed a hybrid search strategy [21] that involves conducting a search string-based database search on a specific digital library (Scopus) and then complementing the set of identified papers with backward and forward snowballing (using Google Scholar) [27], allowing identifying studies indexed in other digital libraries. We intentionally refrained from using various specific libraries (such as Springer Link), but focused on meta search engines for various reasons. Foremost, the topic addressed cannot be assigned to single specific research communities and, thus, publication venues, but is a rather broad, multidisciplinary field of research at the intersection to various software engineering sub-disciplines. Please note that the employed strategy generally tends to provide similar results as when conducting searches on several digital libraries, as shown in [21], while in our case, we would have need strong adaptations of the search strings according to the technical particularities of the single libraries and according to the methodological particularities of the communities behind the venues. The hybrid approach allows, in contrast, a rather simplified string supporting reproducibility and replicability.

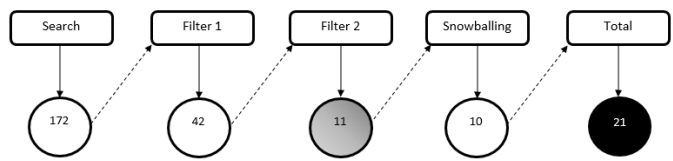

Fig. 1. Filtering process of the papers

We chose, in particular, Scopus, because it claims to be the largest database of titles and abstracts. It is, however, backward and forward snowballing via Google Scholar which we used as an effective way to identify the broader population of studies [21][27].

Thus, the first step was to formulate a search string to conduct the initial database search on Scopus. Our study focuses on the population and context of agile software development and aims at identifying approaches that handle SR (intervention). Thus, we needed keywords for agility and SR.

The defined search string was: *"((Agility OR Agile OR Scrum OR Extreme Programming) AND (Security AND Requirements))"*. It is noteworthy, that the keywords *Scrum* and *Extreme Programming* were included because of the popularity of these agile methods [8]. The search strategy was applied on titles, abstracts and keywords in Scopus in October 2017.

*C. Study Selection*

The primary inclusion criterion was on papers that describe how to handle SR in the context of agile projects. When several papers reported the same study (e.g., in the case of journal extensions), only the most recent one was included. When multiple studies were reported in the same paper, each relevant study was considered separately. The exclusion criteria applied for filtering the papers are shown in Table 1.

TABLE 1. EXCLUSION CRITERIA.

| Criteria | Description |
|---|---|
| EC1 | Papers that do not have information on how they handle SR in the context of agile projects |
| EC2 | Papers not written in English |
| EC3 | Grey literature, including white papers, theses, and papers that were not peer reviewed |
| EC4 | Papers that are only available in the form of abstracts/posters and presentations |

The first step consisted of searching for papers using the search string in the digital library selected for this study (Scopus). The search returned 172 papers, as shown in Fig. 1. In the second step (Filter 1), the first filtering took place. In this step, we applied the exclusion criteria (EC1, EC2, EC3 and EC4). Regarding EC1, at this step we excluded the papers that clearly didn't have information on how to handle SR in agile projects. As a result, we reduced the set of candidate papers to 42. All exclusions were peer reviewed and agreed by an independent researcher.

In the third step, we applied a second filter (Filter2). We filtered papers by reading the titles, abstracts and selected paper parts (when necessary) while applying the inclusion and exclusion criteria (in particular, reviewing EC1 against the inclusion criterion). This step left us with 11 papers representing the result of the database search. All exclusions and the final set of included papers were peer reviewed by an independent researcher. In case of divergence, a third researcher was involved and a discussion was held to reach consensus.

Finally, in the fourth step we applied backward and forward snowballing iteratively following the snowballing guidelines in [27]. In total, 5 backward snowballing and 4 forward snowballing iterations were applied (order: BS1, BS2, FS1, BS3, BS4, FS2, FS3, BS5, FS4). The 5 backward snowballing iterations involved analyzing 384 (183 + 119 + 27 + 20 + 35) papers (including duplicates) and allowed identifying 8 additional papers to be included. The 4 forward snowballing iterations involved analyzing 289 (282 + 1 + 5 + 1) papers (including duplicates) and allowed identifying 2 additional papers to be included. The whole snowballing process was peer reviewed. A spreadsheet with further details on the filtering and snowballing process can be found online[2].

*D. Data Extraction and Classification Scheme*

The information extracted from each of the 21 selected papers and the classification schemes describing the different categories are presented in Table 2.

TABLE 2. DATA EXTRACTION FORM.

| Information | Description |
|---|---|
| Study Metadata | Includes the paper title and information on the authors, venue, and year of publication. |
| Approach | Name of the approach. |
| Description | Short description of the approach. |
| Agile Method (RQ1) | Agile method in which the approach was proposed or applied. |
| RE Phase (RQ2) | RE phase that the approach addresses. To enable classifying the identified approaches independently of their particularities, we defined generic and high-level RE phases following the ReqMan process framework (e.g., elicitation, analysis, specification, validation) according to which any RE approach can be (re-)structured to illustrate the primary purpose of the included practices. |
| How SR are Handled (RQ3) | How solution options are typically built. Based on the approach description we applied the constant comparative method [24] to create categories. |
| Research Type Facet (RQ4) | Classification of research type facets according to Wieringa *et al*. [26], including the following categories: *evaluation research, solution proposal, philosophical paper, opinion paper,* or *experience paper*. |
| Empirical Evaluation (RQ5) | Classification of the empirical strategy, according to Wohlin *et al*. [28], including the following categories: *experiment, case study,* or *survey*. |
| Limitations (RQ6) | Limitations or problems reported within the paper. |

IV. SYSTEMATIC MAPPING RESULTS

This section presents the results of the mapping study. First, we provide an overview of the included papers and the identified approaches. The following subsections present the answers to our research questions based on the information extracted from the included studies.

*A. Overview of the included papers and approaches*

Overall, we identified 21 papers involving 34 different authors. The identified approaches and a short description can be found in Table 3. The full references of the list of selected papers are presented in Appendix A. Regarding the years of publication, the papers presenting the approaches range from 2005 to 2017. While analyzing the distribution of paper throughout the years, there seems to be a growing interest in the area with several papers being published recently. Most of the publications (17) are conference and workshop papers and only 4 papers have been published in journals.

---

[2] https://hrguarinv.github.io/SEAA2018/

TABLE 3. IDENTIFIED APPROACHES ORDERED BY YEAR.

| Id | Year | Description |
|----|------|-------------|
| P12 | 2005 | Provides an example to identify and classify security objects that can be added seamlessly to agile methods. |
| P13 |  | Extends agile practices to deal with security. Abuser story is the artifact discussed. |
| P11 | 2006 | Proposes a way of extending XP practices. Abuser stories and security functionalities are proposed. |
| P14 |  | Presents suggestions on how agile methods are expanded to deal with security. User and misuse stories are discussed. |
| P10 | 2010 | Explains a secure approach based on XP. Techniques such as misuse case and attack tree are proposed in early stages. |
| P2 | 2011 | Proposes a conceptual agile security framework using a hybrid technique for requirements elicitation. |
| P9 |  | Proposes the security backlog element and integrates this feature into the Scrum method. |
| P8 | 2012 | Presents a NORMAP methodology that identify, link and model Agile Loose Cases (ALC), Agile Use Cases (AUC) and Agile Choose Cases (ACC). |
| P15 |  | Presents NORMATIC, a Java-based tool for modeling non-functional requirements (NFR) for agile processes. |
| P18 |  | Models NFR and their potential solutions in a visual environment. |
| P16 | 2013 | Proposes an enhanced version of Scrum to accommodate security analysis and design activities within the method. |
| P1 | 2014 | Evaluates previous research about the security backlog. |
| P6 |  | Proposes the use of security assurance cases to maintain a global view of the security claims in Scrum. |
| P7 |  | Proposes a way to map security activities into the DSDM. A case based reasoning technique is used for maintaining security requirement repository. |
| P17 |  | Proposes a method for security reassurance into the agile methods. Abuser story is proposed to elicit requirements. |
| P4 | 2015 | Proposes to add specific security techniques into Scrum, e.g., assets extraction, security user stories, among others. |
| P5 |  | Presents modifications (e.g., roles, tasks and artifacts) to Scrum to achieve the security levels of VAHTI framework. |
| P19 |  | Proposes NERV: Non-functional Requirements Elicitation, Reasoning, and Validation in agile processes. |
| P3 | 2017 | Maps security NFR and introduces them in the beginning phase of agile methods. |
| P20 |  | Presents the documentation and challenges of NFR in companies utilizing agile methods. |
| P21 |  | Presents elicitation guidelines for NFR in agile methods. It proposes a NFR card. |

*B. (RQ1) In the context of which agile methods have the approaches been proposed or applied?*

Most of the identified approaches have been proposed or applied in the context of Scrum (9 out of 21 studies). Other agile methods comprise XP (2 studies), Feature Driven Development (1 study) and Dynamic Systems Development Method (1 study). It might be interesting to observe that a large number of authors propose generic approaches (8 out of 21) that, according to them, can be rationally adapted to any agile method.

*C. (RQ2) Which RE phases do the approaches address?*

The majority of the selected papers concern requirements specification practices (12 out of 21). Furthermore, we found eight papers about requirements elicitation, three concerning requirement analysis [P10][P15][P18] and only one about requirement validation [P19]. It is important to note that some papers deal with topics comprising more than one phase (e.g., [P8] [P17]). However, a focus lies on both the conceptual specification and document layer and the analytical layer while validation and general quality assurance topics seem currently out of scope.

*D. (RQ3) How are SR handled in each approach?*

During the analysis of the approaches, it was possible to identify and group similar types of solution options that would allow us to better understand how SR are handled in agile contexts. These solution option types were identified by applying the constant comparative method [24], to compare the description of each approach against a cataloged list. At all, we identified five different types, which are described hereafter.

*Modifying agile methods:* One of the most common ways to handle SR in the context of agile methods is to propose modifications to them, e.g., by integrating a new feature such as a security backlog [P9] or including new activities such as vulnerability analyses [P4]. This type of solution option was employed in 7 out of the 21 papers [P1][P3][P4][P5][P7][P9][P16].

*Introducing new artifacts:* Five of the identified approaches involve creating new artifacts, such as extending the concept of user story to abuser story [P10][P12][P13][P14][P17]. User stories are often written from the perspective of an end user. Abuser story artifacts, on the other hand, represent the point of view of a malicious adversary (in analogy to misuse cases in rather plan-driven RE approaches).

*Introducing guidelines to handle security issues*: In this case, seven of the approaches involve specific guidelines to be followed by stakeholders [P6][P8][P11][P18][P19][P20][P21]. This kind of solution option can be seen as a systematic way of doing activities that are already comprised in existing agile methods, e.g., [P21] presents an agile non-functional elicitation process to be executed step by step.

*Proposing a conceptual framework*: Only one approach presents a conceptual framework for handling SR that could be adapted and applied to any agile method [P2]. Contrary to the previous solution type, such frameworks allow to be flexible in its usage, e.g., it implements security practices in agile development and adopts additional features proposed by other researchers.

*Providing tool support*: Again, only one approach proposed a software tool in order to deal with SR in the context of agile methods [P15]. In this case, a conceptual approach was implemented as a software solution (a java-based simulation tool was developed for modeling its initiatives).

*E. (RQ4) What is the research type facets of the approaches?*

Figure 2 shows the distribution of the research type facets of the papers per year. It is possible to observe that most of the papers (12 out of 21) concern solution proposals. It is noteworthy that there seems to be a trend of gradually moving from opinion papers over solution proposals to very first evaluation papers. The latter is, however, at its beginning.

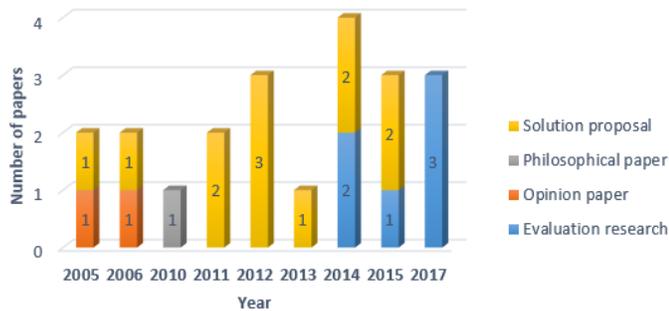

Fig. 2. Distribution of research type per year.

### F. (RQ5) Which kind of empirical evaluations have been performed?

When analyzing the empirical evaluations conducted within the studies (Fig. 3), it was possible to identify 10 papers that have performed empirical evaluations (9 case studies and 1 controlled experiment). It is important to note that 4 papers from this set complemented their solution proposals by simplified case studies, therefore they were not classified as evaluation research. On the other hand, while it is noteworthy that 11 studies did not contain any type of empirical evaluation, a majority of those who did favored in-vivo studies over in-vitro studies. This is in tune with our expectation that many challenges in integrating SR in agile practices can be observed in their natural environment only. Most of the recent papers include some empirical evaluation method such as controlled experiment [P7] and cases studies [P3][P20][P21].

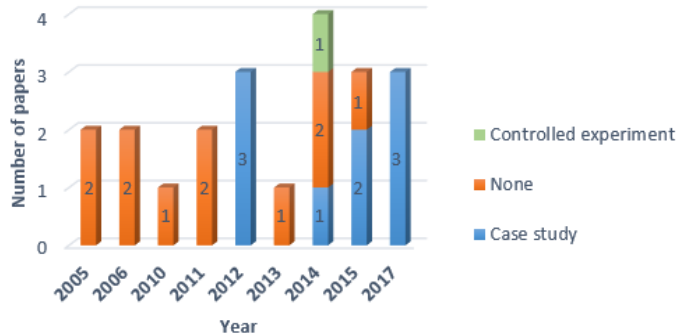

Fig. 3. Distribution of empirical evaluation types per year.

### G. (RQ6) What are the limitations faced by the identified approaches?

Some papers explicitly report limitations of the agile SR approaches concerning *environment*, *people*, or *effort or resources*. A brief overview of the identified limitations is summarized below.

*Environment.* As the Scrum release cycle is too short, there seems to be not enough time for the development team to address SR for each release. In [P1][P4], the authors state this as an important limitation when using Scrum.

*People.* In [P1][P12] the authors state that knowledge monopolies in a Scrum team are the main factors causing an absence of documentation during requirement planning. This happens due to the team players not having enough skill in regards to security. Furthermore, [P1] reports that, although the project manager played a major role in Scrum, their lack of security awareness and pressure to complete the project within a minimal amount of time affected the entire process.

*Effort and resource.* The lack of guidelines in the collection of SR is reported as a limitation [P1][P9]. Additionally, as agile methods are essentially iterative, the periodic security reassurance of a software increases the cost of security assurance [P17]. Furthermore, several authors [P1][P2][P5][P7][P9][P11] propose to create a new role for security concerns or to provide security training to developers, which means additional resources and efforts.

## V. SUMMARY AND DISCUSSION

In this section, we provide a summary of our findings and discuss aspects we personally found particularly important during our research.

Our SM allowed identifying and characterizing 21 different approaches to handle security issues in agile contexts. We invested significant effort in searching for these approaches complementing the database search with iterative forward and backward snowballing. It is noteworthy that only 4 [P8][P13][P15][P18] out of the 21 approaches listed in this mapping study have been included in the secondary studies representing our related work (*cf*. Section II). This highlights the importance of publishing our mapping study to share the results with the community.

We were able to group these approaches into 5 types of solution options. Fig. 4 presents a visualization of part of the SM through a bubble chart, confronting the type of solution options against the RE phases and empirical evaluation types. When analyzing the types of solution options there seems to be a gap concerning tool support, given that only one of the approaches aims at proving such support.

The information on the RE phases helps understanding the purpose of the introduced approaches. It is possible to observe that most of the solution approaches concern requirements specification and elicitation. Thus, there also seems to be a gap concerning security requirements verification and validation in agile contexts.

Regarding the empirical evaluations, it is evident that there are additional gaps to cover, given that there is the lack of empirically evaluated studies. This means we have relatively little evidence about the feasibility of the presented approaches. Given this scenario and subtle but important differences between the approaches, further evaluation research should be conducted in order to better understand the situations in which they can be safely applied and their limitations.

Indeed, little is known about the limitations when using SR in agile contexts (e.g., concerning environment, people and efforts and resources). Nevertheless, we believe that the information provided on limitations in our mapping is still valuable to practitioners, given that many organizations are being attracted by agile methods, but might not be aware of the risks related to security issues.

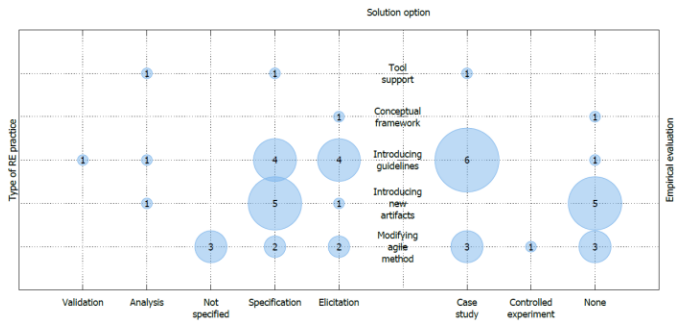

Fig. 4. State of reported evidence: Distribution of type of solution options against the RE phases and empirical evaluation types.

## VI. THREATS TO VALIDITY

In this section, we evaluate our findings and critically review our study regarding its threats to validity. We list the possible threats and procedures we took to mitigate those issues hereafter.

*Reliability:* As any literature study, our study suffers from potential incompleteness of the search results. To address this threat, we used a hybrid search strategy that complements database search (on Scopus) with iterative backward and forward snowballing (using Google Scholar), which resulted in analyzing an extensive list of candidate publications to be included. Hence, we believe that the likelihood of still missing an approach is very low. Another threat that could affect reliability is general publication bias, i.e., positive results are more likely published than negatives ones. Unfortunately, we had to accept this threat. Nevertheless, we analyzed the limitations stated by the authors within the papers.

*Internal validity:* We followed the guidelines provided by Kitchenham and Charters [14]. During paper selection, all exclusions and the final set of included papers were peer reviewed by an independent researcher. In case of divergence, a third researcher was involved and a discussion was held to reach consensus. The snowballing process and the data extraction (including classification) was also peer reviewed. The data and details on the selection process, including intermediate and auditable results, are available online[2].

*External validity*: The external validity concerns the generalizability of the results. While effort has been invested to produce complete, valid and auditable results, our mapping protocol (*cf.* Section III) can be used for further updates and/or replication studies to reinforce them.

## VII. CONCLUDING REMARKS

This paper presents the results of a SM study on SR engineering in the context of agile methods. In this review, a total of 21 studies, published between 2005 and 2017, were identified and analyzed. We classified those studies in terms of the agile methods, RE phases, type of solution approach, research type, and the conducted empirical evaluation. We also analyzed the limitations reported within the studies when handling SR in agile contexts.

Our results show that most approaches relate to the Scrum method. Regarding the type of RE phases, specification and elicitation are the most addressed ones. Concerning the types of solution option, the approaches typically involve modifying agile methods, introducing new artifacts (e.g., extending the concept of user story to abuser story), or introducing guidelines to handle security issues. Other two types of solution options involve proposing a conceptual framework and providing tool support. However, few approaches employ these solution options. When analyzing the type of research and of empirical studies, we observed a lack of evaluation research. Finally, we identified limitations of using these approaches related to environment, people, effort and resources.

Hence, the main identified gaps of our mapping study concern lack of research on SR verification and validation, limited tool support, and a general lack of empirical evaluations. Nevertheless, given its practical relevance, the field seems to be evolving and raising interest and the herein provided mapping study could help to lay the foundations for several promising avenues of future research.